# Triplet-sensitization by lead halide perovskite thin films for near-infrared-to-visible upconversion

*Lea Nienhaus,[‡,&,*] Juan-Pablo Correa-Baena,[†] Sarah Wieghold,[†,&] Markus Einzinger,[¶] Ting-An Lin,[¶] Katherine E. Shulenberger,[‡] Nathan D. Klein,[‡] Mengfei Wu,[¶] Vladimir Bulović,[¶] Tonio Buonassisi,[†] Marc A. Baldo,[¶,*] and Moungi G. Bawendi[‡,*]*

[‡]Department of Chemistry, [¶]Department of Electrical Engineering and Computer Science, [†]Department of Mechanical Engineering, Massachusetts Institute of Technology, Cambridge, MA 02139

*corresponding authors: nienhaus@chem.fsu.edu, baldo@mit.edu and mgb@mit.edu

**ABSTRACT**

Lead halide-based perovskite thin films have attracted great attention due to the explosive increase in perovskite solar cell efficiencies. The same optoelectronic properties that make perovskites ideal absorber materials in solar cells are also beneficial in other light-harvesting applications and make them prime candidates as triplet sensitizers in upconversion *via* triplet-triplet annihilation in rubrene. In this contribution, we take advantage of long carrier lifetimes and carrier diffusion lengths in perovskite thin films, their high absorption cross sections throughout the visible spectrum, as well as the strong spin-orbit coupling owing to the abundance of heavy atoms to sensitize the upconverter rubrene. Employing bulk perovskite thin films as the absorber layer and spin-mixer in inorganic/organic heterojunction upconversion devices allows us to forego the additional tunneling barrier owing from the passivating ligands required for colloidal sensitizers. Our bilayer device exhibits an upconversion efficiency in excess of 3% under 785 nm illumination.

Photon upconversion describes the generation of high-energy photons by combining two or more low-energy photons. This conversion of low-energy photons to high-energy photons provides a means to overcome the Shockley-Queisser limit in single-junction solar cells[1] by sub-bandgap sensitization of the absorber layer,[2–4] and shows promise in near-infrared (NIR) sensitization of low-cost silicon-based devices for applications such as biological imaging,[5,6] 3-dimensional displays[7] and photocatalysis.[6] In this work, we focus on upconversion *via* diffusion-mediated triplet-triplet annihilation (TTA) in organic semiconductors.[8,9] Unlike other upconversion processes such as such as second harmonic generation in nonlinear crystals or upconversion using lanthanides ions,[6] TTA can be effective even at sub-solar photon fluxes since the energy is stored in long-lived triplet states.[10,11]

In the TTA process investigated here, two triplet states on neighboring annihilator molecules (rubrene) generate a higher-lying emissive singlet state by electronic coupling. Direct optical excitation into a spin-triplet state would violate the selection rules, therefore these are "spin-forbidden" and unable to radiatively couple to the ground state. We require sensitizers to convert optically excited spin-singlet states to optically inactive triplet states in rubrene, which are then transferred to the triplet state of rubrene *via* a spin-allowed exchange-mediated (Dexter)[12] energy transfer process.

Historically, sensitization has been achieved by using metal-organic complexes containing heavy metal atoms, which facilitate intersystem crossing (ISC) due to strong spin-orbit coupling. However, this approach has the disadvantage of large exchange energies, which can result in energy losses up to 300 meV.[10,13,14] More recently, we showed lead sulfide (PbS) nanocrystals (NCs) to be efficient triplet sensitizers.[11,15–19] This system has the major advantage of exchange energies on the order of kT (~25meV), resulting in negligible energy loss during the singlet-triplet conversion. Strong spin-orbit coupling in PbS NCs renders the spin quantum number meaningless, and only the total angular momentum quantum number accurately describes the states. As a result, the spin of the optically excited exciton rapidly dephases and its wave function shows both singlet and triplet character.[20,21]

However, current PbS NC based upconversion devices exhibit poor exciton diffusion lengths which leads to inefficient exciton transport within the PbS NC films. As a result, the PbS NC layer thickness is limited to one or two monolayers, causing these devices to suffer from low NIR absorption.[17,19] To circumvent the issue of poor exciton transport within large-scale NC arrays, we seek to replace the "bulk" NC array with a true bulk semiconductor material which shows long carrier diffusion lengths and also can undergo efficient energy transfer to rubrene.

Lead halide perovskite thin films have shown great promise as absorber layers in photovoltaics. Commonly used in perovskite solar cells (PSCs), this technology has made impressive progress in just a few years with efficiencies climbing from 3.8%[22] in 2009 to a certified 22.7%[23] in 2017. The general formula of the commonly used perovskite is $ABX_3$, containing an organic cation $A$, such as methylammonium (MA) or formamidinium (FA),[22,24] a divalent metal $B$, such as lead or tin,[25] and a halogen $X$, such as bromine or iodine. The impressive performances achieved to date have been attributed to exceptional material properties such as low non-radiative recombination,[26] low exciton binding energies,[27–29] high light absorption over the visible spectrum, and charge carrier diffusion lengths in the micrometer range due to long carrier lifetimes.[30–32] All of these characteristics allow for efficient extraction of charges at the interface of the electron- and hole-transport layers in photovoltaic devices.

Unlike the PbS NCs investigated previously,[15,16,19] which exhibit long-lived excitonic (bound electron-hole pairs) behavior, the initially-excited bound excitons in MA lead triiodide (MAPbI$_3$) perovskite films commonly dissociate within picoseconds into unbound and highly mobile charge carriers.[33] Due to the abundance of heavy atoms (Pb, I) in lead halide perovskites, strong spin-orbit coupling is expected in the perovskite film and the electron and hole can both rapidly spin-mix.[34,35] Hence, injection of free electrons and holes into the upconverting organic semiconductor can provide a new avenue for sensitization of rubrene, and may allow us to move away from the necessity of efficient singlet-to-triplet exciton converters.[36]

Here, we investigate the upconversion in rubrene sensitized by $MA_{0.15}FA_{0.85}PbI_3$ thin films. The rubrene layer is doped with 1% dibenzotetraphenylperiflanthene (DBP). This host-guest/annihilator-emitter approach is adopted from organic light-emitting diodes and has been shown to enhance the quantum yield (QY) of the rubrene film, as singlets generated by the upconversion process in rubrene are harvested *via* Förster resonance energy transfer (FRET) to the emitter DBP, outcompeting singlet fission.[15]

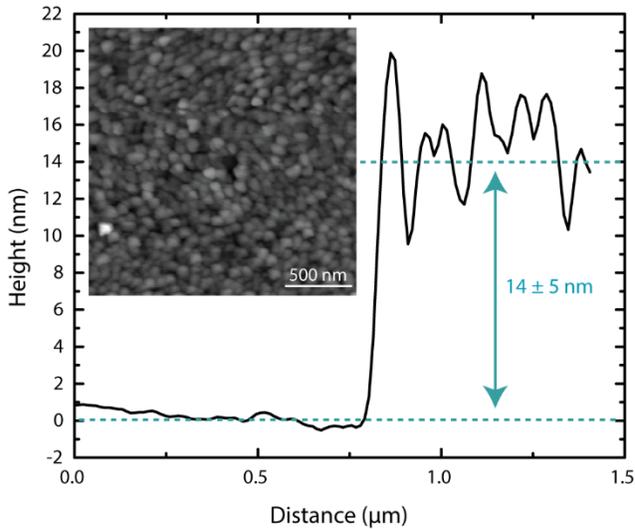

*Figure 1: AFM height profile of the $MA_{0.15}FA_{0.85}PbI_3$ thin film. The inset shows the AFM image of the perovskite film, with a grain size distribution of 50 ± 20 nm.*

Atomic force microscopy (AFM) is used to determine the morphology and thickness of the perovskite film. Figure 1 shows a line profile across a boundary of the perovskite film and the underlying substrate yielding a perovskite film thickness of ~14 nm. The inset shows the morphology of the perovskite film, highlighting the small grain size of ~50 nm (compare also SI Figure S1 & S2). This grain size has previously been shown to be large enough to have a true "bulk" behavior of the excited charge carriers and avoid any quantum confinement effects which may change the underlying carrier dynamics.[37] Quantum confinement effects are known to increase the bandgap of semiconductors. We further confirm the lack of energetic confinement and thus, the bulk-like behavior of our $MA_{0.15}FA_{0.85}PbI_3$ thin films by steady-state photoluminescence (PL) spectroscopy (Figure 2a, brown), where we see the peak emission at the expected value of ~1.55 eV (800 nm) for bulk $MA_{0.15}FA_{0.85}PbI_3$.[38]

The absorbance of the neat $MA_{0.15}FA_{0.85}PbI_3$ thin film (black line) and of the bilayer upconversion device (gray), as well as the emission at 610 nm of the upconverted light (red) are shown in Figure 2a. The inset shows the upconverted light created in the upconversion device under 785 nm illumination, observed as the orange line running horizontally through the device. Figure 2b depicts the upconversion device structure investigated here consisting of the 15 nm thin film of $MA_{0.15}FA_{0.85}PbI_3$, with a subsequent layer of rubrene/1% DBP spin coated from a 10 mg/ml stock solution.

In Figure 2c the band diagrams of the $MA_{0.15}FA_{0.85}PbI_3$ perovskite and rubrene are shown.[16,38] Two triplet sensitization mechanisms are possible: i) direct triplet exciton transfer (TET) of a bound triplet exciton as reported previously for quantum confined perovskite nanocrystals[39] or as in the case of PbS sensitized upconversion[15,16] and ii) sequential electron and hole transfer resulting in a bound triplet on the rubrene molecule.

For sensitization *via* direct TET from the initially created band-edge excitons to occur at the interface, the combined rate of exciton diffusion to the interface and TET must outcompete the exciton dissociation rate. However, as discussed previously, lead halide perovskites exhibit binding energies below $kT$ and the initially optically excited band-edge excitons generally rapidly dissociate within picoseconds. Hence, TET from these band-edge excitons is unlikely to be the dominant mechanism of triplet sensitization at room temperature by the bulk perovskite investigated here. However, shallow traps may result in localized "long-lived" sub-bandgap excitons which could to transfer to the rubrene layer within their lifetime.[40]

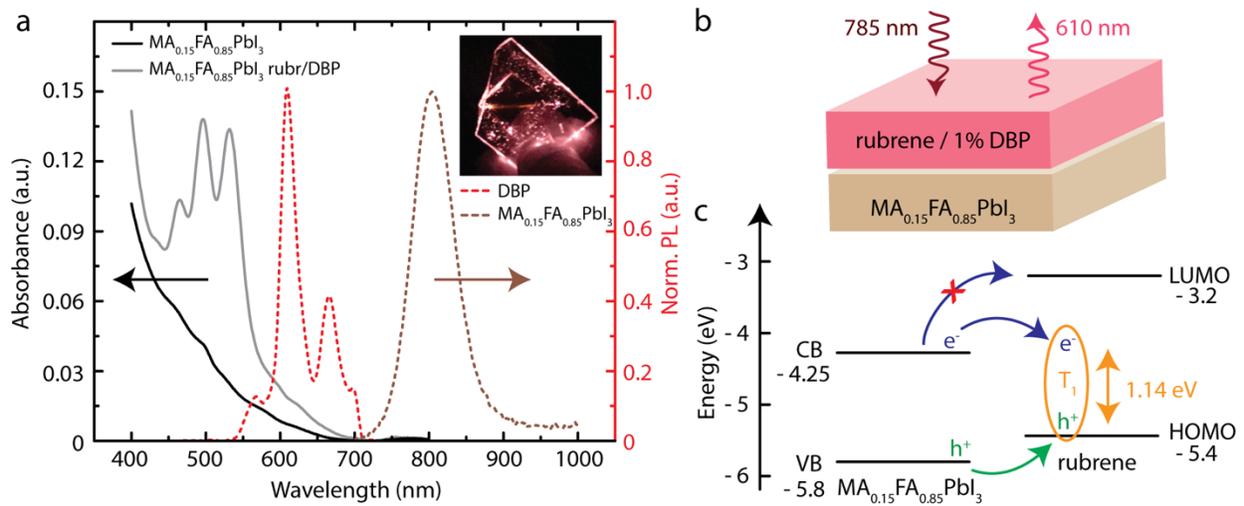

Figure 2: (a) Normalized absorbance spectra of a neat $MA_{0.15}FA_{0.85}PbI_3$ perovskite film (black), $MA_{0.15}FA_{0.85}PbI_3$/rubrene/1%DBP bilayer device (gray), the upconverted emission stemming from rubrene/DBP (red) and the NIR emission of the $MA_{0.15}FA_{0.85}PbI_3$ perovskite (brown). The inset shows the device under 785 nm illumination, and the upconverted light is seen as the orange line. (b) Schematic of the bilayer upconversion device, which emits 610 nm light upon 785 nm excitation. (c) Band diagram of $MA_{0.15}FA_{0.85}PbI_3$ and rubrene, showing the energetically favored hole transport from the perovskite VB to the rubrene HOMO. The electron cannot directly transfer from the perovskite CB to the rubrene LUMO due to an energy barrier of ~1eV.

The energetic alignment indicates that hole transfer from the perovskite valence band (VB) to the rubrene highest occupied molecular orbital (HOMO) can be very efficient due to the driving energy of ~0.4 eV. An unbound electron on the other hand, cannot transfer directly into the lowest unoccupied molecular orbital (LUMO) of rubrene due to the ~1 eV energy gap between the perovskite conduction band (CB) and the rubrene LUMO, despite significant band bending reported by Ji *et al.*[41] However, with a hole residing on the rubrene molecule, it may be possible to transfer the electron directly into the bound triplet state ($T_1$) through a mediating charge-transfer state (CT) at the interface.

To further investigate the upconversion process, as well as determine the timescale of triplet sensitization, we turn to time-resolved transient PL spectroscopy. Figure 3a shows the time-resolved NIR PL of a neat $MA_{0.15}FA_{0.85}PbI_3$ film (black trace), and of the bilayer upconversion device (red trace). Both decay traces show highly multiexponential dynamics, indicating a broad distribution of emission rates. To avoid overfitting, we define the PL lifetime of the neat $MA_{0.15}FA_{0.85}PbI_3$ as the time in which the population decays to 1/e ($\tau_{1/e}$ = 50 ns). The very long tail ($\tau_l$ = 8 µs) is attributed to delayed fluorescence stemming from trapped carriers.[42] Interestingly, we observe that the overall emission intensity of the $MA_{0.15}FA_{0.85}PbI_3$ perovskite film is not notably quenched as previously seen in PbS NC-based devices,[15,16] despite seeing a decrease in the PL lifetime. This residual NIR PL intensity can likely be attributed to carriers that radiatively recombine in the "bulk" of the perovskite film prior to reaching the extraction point at the $MA_{0.15}FA_{0.85}PbI_3$/rubrene interface. In addition, the organic molecules can passivate dangling bonds at the surface of the perovskite lattice. This reduces non-radiative recombination occurring at surface defects in the $MA_{0.15}FA_{0.85}PbI_3$ perovskite film, thus increasing the overall PLQY.[43,44]

To extract the dynamics of energy transfer process at the $MA_{0.15}FA_{0.85}PbI_3$/rubrene interface, we apply our previously introduced extraction method,[15,16] where we subtract the residual NIR PL dynamics of the neat $MA_{0.15}FA_{0.85}PbI_3$ film from the dynamics of the bilayer device. This method relies on the basis of an active and an inactive population. The active population couples strongly to the rubrene and is quenched by the added upconverting layer, while the inactive population does not interact with the rubrene. This can be due to the morphology of the rubrene layer (*e.g.* pinholes in the rubrene film), or be a result of carriers which radiatively recombine within the thin perovskite film prior to reaching the inorganic/organic interface. The transfer kinetics obtained by this method are seen in Figure 3a as green dots. We find a characteristic decay time of $\tau_2 = 15 \pm 5$ ns (dark blue), highlighted in the inset in Figure 3a, which shows a magnification of the early time dynamics. The observed subsequent rise and slow decay in the extracted transfer dynamics (green dots) at later times can be attributed to back transfer of singlet excitons created after diffusion-mediated TTA.[16]

The extracted transfer time is slower than reported previously for quenching by hole- or electron-acceptors[30,45] and may indicate the presence of a long-lived state which mediates the energy transfer process as observed previously in triplet sensitization by PbS NCs[46,47] or could stem from an energy barrier for the electron transfer process due to the reported band bending at the perovskite/rubrene interface.[41]

To further investigate the upconversion mechanism, we investigate the dynamics of the upconverted emission, which correlate to a convolution of the rate of energy transfer, diffusion-mediated TTA in rubrene and the emission from DBP following FRET. Figure 3b highlights the transient PL spectroscopy performed on a bilayer device when monitoring the DBP emission. The inset shows an enlargement of the PL dynamics at a short time scale. We observe that the long-lived triplets ($\tau_{triplet} > 12$ μs) have not fully decayed at 25 μs, which results in a buildup of triplet excitons in the rubrene film that do not decay between pulses. An increase in the triplet population has been shown to increase the rate of TTA, which results in an apparent reduction in the extracted characteristic time for TTA, as well as the extracted triplet lifetime.[15,16] The extracted rise time $\tau_{rise} = 24 \pm 3$ ns of the upconverted emission therefore portrays only a lower bound for the rate of the total upconversion process. Interestingly, the peak emission does not occur until t = 3 µs, and a second slower rise is observed leading up to this time. We attribute this to additional delayed TTA caused by slow triplet diffusion and by delayed triplet generation by thermally detrapped carriers, which can occur in addition to the prompt transfer of free carriers.

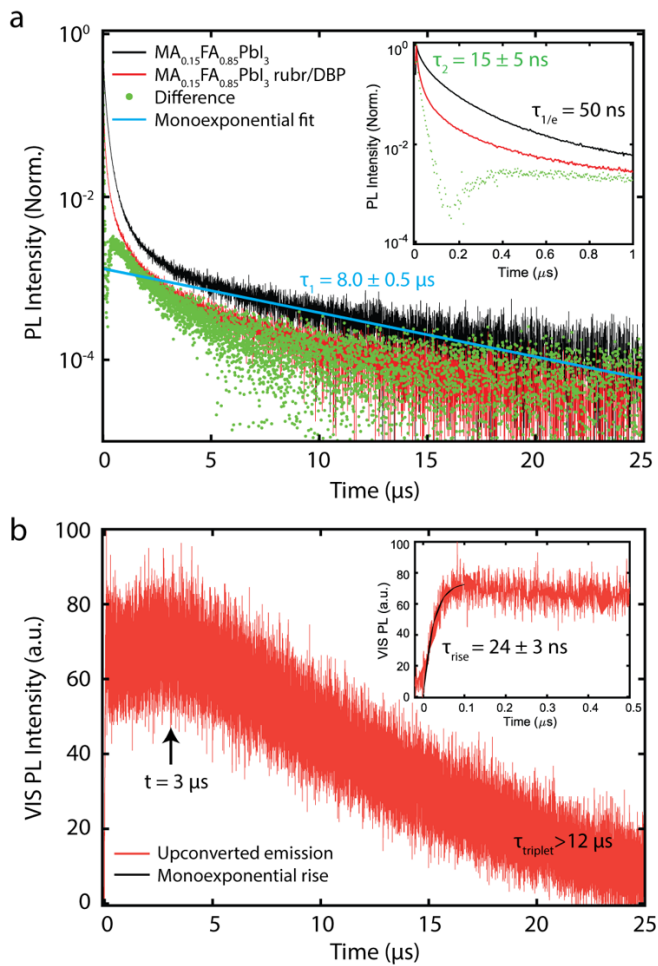

Figure 3: (a) Time-resolved PL dynamics of the NIR emission from neat $MA_{0.15}FA_{0.85}PbI_3$ (black), the quenched NIR dynamics of $MA_{0.15}FA_{0.85}PbI_3$ in the bilayer device (red) under 785 nm excitation. The trapped carrier lifetime ($\tau_1$ = 8 µs) is extracted from the long-time dynamics (light blue). The extracted dynamics (green) yield an extracted quenching rate of $\tau_2 = 15 \pm 5$ ns. (b) Dynamics of the upconverted light emitted from the bilayer device. The rise time gives an upper bound for the triplet sensitization rate ($\tau_{rise} = 24 \pm 3$ ns). The insets show an enlargement of the early-time dynamics.

However, as both bound triplet excitons and free carriers (holes) can be present in the rubrene layer, triplets may be annihilated not only by TTA, but also through triplet-charge annihilation (TCA). In the TCA process, a bound triplet exciton collides with an adjacent single charge and either scatters it, or the collision dissociates the bound triplet exciton into free charges. Therefore, this latter mechanism creates more freely diffusing charges in the film which can undergo TCA, effectively reducing the achievable upconverted PL.[35,48] To further elucidate roles of TCA and TTA we investigate the magnetic field effect (MFE) of the upconverted light, as well as the power dependence of the emission, both of which exhibit features unique to TTA.

The theory of MFEs in molecular crystals was first developed by Merrifield in 1968.[49,50] When two triplet excitons collide, one of nine possible triplet-triplet pair states is formed. Only those with singlet character can effectively couple to the singlet state. The external magnetic field can modify the number of states with singlet character. At zero field only three of the nine states have singlet character, and as the field is increased, additional states acquire singlet character, thus increasing the rate of TTA. At high fields, the singlet character is localized on only two pair states, and therefore the yield of TTA is reduced, resulting in the negative MFE typical of TTA. Figure 4a shows the $MFE_{UC}$ of the upconverted emission from a $MA_{0.15}FA_{0.85}PbI_3$/rubrene/1%DBP bilayer device, which shows the typical signature of TTA (red hexagons). The $MFE_{PL}$ of a neat $MA_{0.15}FA_{0.85}PbI_3$ film under 532 nm excitation is shown as black squares, and despite reports of negative MFEs (up to 2.5% on photocurrent at 300mT) in similar perovskite thin films[35,51] we are not able to unambiguously detect an effect by the applied magnetic field.

Furthermore, the power dependency of the $MA_{0.15}FA_{0.85}PbI_3$ PL (~800 nm) and of the upconverted emission of a $MA_{0.15}FA_{0.85}PbI_3$/rubrene/1%DBP bilayer device, under 785 nm pulsed excitation at 10 MHz is highlighted in the log-log plot in Figure 4b (compare also to SI Figure S3). The $MA_{0.15}FA_{0.85}PbI_3$ PL (black squares) shows a slightly super-linear power dependency, with a slope of 1.1 (gray line) indicating the emission primarily stems from monomolecular recombination in the perovskite thin film, consistent with previous reports of shallow-trap-assisted recombination.[29,52,53]

The upconverted emission (red hexagons) shows the transition of the slope from quadratic (slope = 2) to linear (slope = 1), which is typical of bimolecular TTA.[54] In the weak annihilation regime (lower powers) triplets decay primarily *via* quasi first-order kinetics and the upconverted PL intensity is quadratic with excitation power (white background). At higher powers (strong annihilation regime), the triplets decay primarily through bimolecular TTA, resulting in a linear dependence of the upconversion efficiency and the excitation power (green background).

As a result, the upconversion efficiency is dependent on the excitation power and the maximal upconversion efficiency is obtained in the linear regime. In the $MA_{0.15}FA_{0.85}PbI_3$/rubrene/1%DBP bilayer devices this quadratic-to-linear transition occurs at 500 mW/cm$^2$ (dashed green line) average incident power under pulsed excitation, nearly an order of magnitude lower than the incident power required in previous PbS NC-based upconversion devices,[15] and two-fold less than in interference enhanced upconversion devices.[15]

Finally, we investigate the upconversion efficiency ($\eta_{UC}$) of the bilayer device. $\eta_{UC}$ is defined as the fraction of excited carriers in the sensitizer that are converted to singlet excitons in the annihilator,[6] which is the product of the energy transfer efficiency from sensitizer to annihilator ($\eta_{ET}$) and the TTA efficiency of the annihilator ($\eta_{TTA}$). On the other hand, the external quantum efficiency (EQE) is the ratio of number of upconverted visible photons to number of incident infrared photons. In our bilayer device, this is further equivalent to the product of absorption of

$MA_{0.15}FA_{0.85}PbI_3$ at 785nm excitation ($Abs_{785nm}$), the upconversion efficiency ($\eta_{UC}/2$), and the PLQY of rubrene:1 wt.% DBP film ($\eta_{Rb:DBP}$) at visible excitation, as shown in equation (1):

$$EQE = Abs_{785nm} \cdot \left(\frac{\eta_{UC}}{2}\right) \cdot \eta_{Rb:DBP} \tag{1}$$

Note that $\eta_{UC}$ is divided by 2 to normalize to a maximum efficiency of 100%. Both the EQE of the bilayer device under 785nm excitation and $\eta_{Rb:DBP}$ at 450nm excitation are measured with an integrating sphere,[55] giving an EQE of $(1.5 \pm 0.3) \cdot 10^{-4}$ % and $\eta_{Rb:DBP}$ of $9.3 \pm 0.8$ %.

However, due to the very low absorption of $MA_{0.15}FA_{0.85}PbI_3$ thin films at the excitation wavelength of 785 nm, and parasitic interference effects, it is difficult to accurately measure the percentage of absorbed photons (compare also SI Figure S4). We estimate the upper bound of the absorption to be 0.1%, which yields a conservative estimate for the internal upconversion efficiency of $\eta_{UC} = 3.1\%$.

In conclusion, we have been able to show that $MA_{0.15}FA_{0.85}PbI_3$ thin films can function as efficient triplet sensitizers of the annihilator rubrene. The presented results show that TTA is the prevalent process observed in the $MA_{0.15}FA_{0.85}PbI_3$/rubrene/1%DBP bilayer device, with no detectable detrimental effect caused by TCA. The extracted triplet sensitization rate of $\tau = 15$ ns is unusually slow for an electron transfer process and may point to a long-lived intermediate localized bound excitonic state mediating the energy transfer process, or an energy barrier for the electron transfer at the interface stemming from the reported band bending and will be subject to further investigation. Optimization of the energy transfer process, the energetic band alignment in the upconversion device, as well as the device structure provide a means to further improve the device performance.

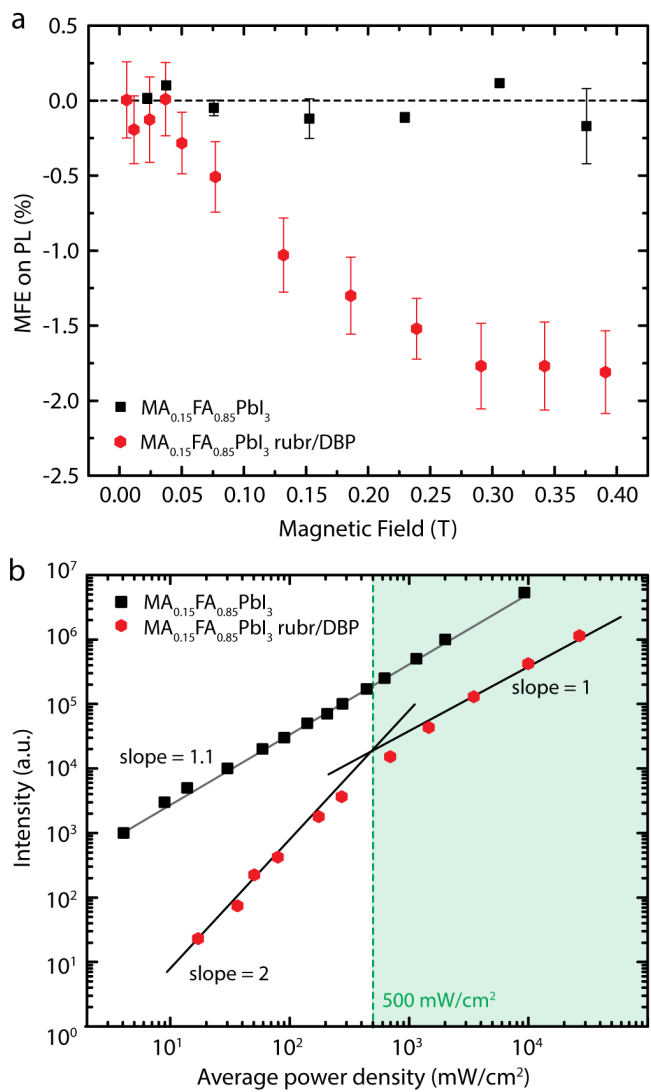

*Figure 4: (a) Change in the upconverted emission (red hexagons) and the NIR $MA_{0.15}FA_{0.85}PbI_3$ emission (black squares) under an applied magnetic field. (b) Power dependency of the NIR emission of the $MA_{0.15}FA_{0.85}PbI_3$ perovskite in a bilayer device, showing a slightly super-linear trend (black squares). The upconverted emission (red hexagons) shows the quadratic-to-linear slope change typical of TTA, with an average incident power density threshold of 500 mW/cm$^2$.*

## ASSOCIATED CONTENT

## Supporting Information

The Supporting Information is available free of charge at DOI:

Experimental section, additional AFM data (Figure S1 & S2), upconverted PL dynamics during the power dependency (Figure S3), and the absorbance spectra of the bilayer device showing the parasitic absorption effects (Figure S4).

## AUTHOR INFORMATION

### Corresponding Author

*Lea Nienhaus: nienhaus@chem.fsu.edu

*Moungi G. Bawendi: mgb@mit.edu

*Marc A. Baldo: baldo@mit.edu

### Present address

&Department of Chemistry and Biochemistry, Florida State University, Tallahassee, FL 32306

### ORCID

Lea Nienhaus: 0000-0003-1412-412X

Juan-Pablo Correa-Baena: 0000-0002-3860-1149

Sarah Wieghold: 0000-0001-6169-3961

Marc A. Baldo: 0000-0003-2201-5257

Moungi G. Bawendi: 0000-0003-2220-4365

### Notes

The authors declare no competing financial interests.

## ACKNOWLEDGEMENT

This work was supported as part of the Center for Excitonics, an Energy Frontier Research Center funded by the US Department of Energy, Office of Science, Office of Basic Energy Sciences under Award Number DE-SC0001088 (MIT). J.P.C.B. is supported by a postdoctoral fellowship awarded by the U.S. Department of Energy, Office of Science, Office of Energy Efficiency and Renewable Energy. S.W. was supported by the National Science Foundation (NSF) under Grant NSF CA No. EEC-1041895. N.D.K. was supported by the National Science Foundation Graduate

Research Fellowship under Grant No. 1122374 and funded by the US Department of Energy, Office of Basic Energy Sciences, Division of Materials Sciences and Engineering under Award No. DE-FG02-07ER46454. This work was supported in part at the Center for Nanoscale Systems (CNS), a member of the National Nanotechnology Coordinated Infrastructure Network (NNCI), which is supported by the National Science Foundation under NSF Award No. 1541959. CNS is part of Harvard University.

**TOC Graphic**

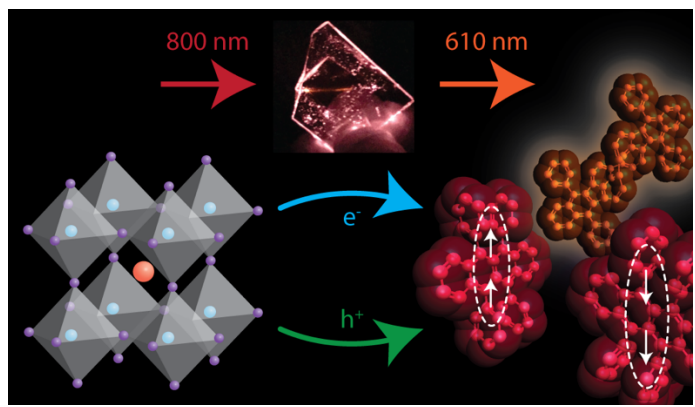